\documentclass{article}
\usepackage{ijcai16}
\usepackage{color}
\usepackage{amsfonts}
\usepackage{float}
\usepackage{amsmath}
\usepackage{graphicx}

\let\stdforall\forall
\renewcommand{\forall}{\mathop{\vphantom{\sum}\mathchoice
  {\vcenter{\hbox{\Large$\stdforall$}}}
  {\vcenter{\hbox{\Large$\stdforall$}}}{\stdforall}{\stdforall}}\displaylimits}
\let\stdexists\exists
\renewcommand{\exists}{\mathop{\vphantom{\sum}\mathchoice
  {\vcenter{\hbox{\Large$\stdexists$}}}
  {\vcenter{\hbox{\Large$\stdexists$}}}{\stdexists}{\stdexists}}\displaylimits}

\usepackage{times}

\title{Towards fully automated protein structure elucidation with NMR spectroscopy}
\author{Piotr Klukowski$^{1}$, Adam Gonczarek$^{1,2}$ \\ 
Faculty of Computer Science and Management, Wroc\l aw University of Science and Technology$^{1}$ \\
Alphamoon Ltd., ul. W\l odkowica 21/3, 50-072 Wroc\l aw, Poland$^{2}$\\
\{piotr.klukowski, adam.gonczarek\}@pwr.edu.pl}

\begin{document}

\maketitle

\begin{abstract}
Nuclear magnetic resonance (NMR) spectroscopy is one of the leading techniques for protein studies. The method features a number of properties, allowing to explain macromolecular interactions mechanistically and resolve structures with atomic resolution. However, due to laborious data analysis, a full potential of NMR spectroscopy remains unexploited. Here we present an approach aiming at automation of two major bottlenecks in the analysis pipeline, namely,  \textit{peak picking} and \textit{chemical shift assignment}. Our approach combines deep learning, non-parametric models and combinatorial optimization, and is able to detect signals of interest in a multidimensional NMR data with high accuracy and match them with atoms in medium-length protein sequences, which is a preliminary step to solve protein spatial structure. 
\end{abstract}

\section{Introduction}

According to Protein Data Bank (PDB), NMR spectroscopy has been used to solve over 12 000 structures of macromolecules. This process usually requires recording about 10-20 volumetric images (3D, 4D), which we refer to as NMR spectra (see Fig. 1a). Each of them encodes a partial information about a protein structure. For now, no robust approach for fully automated analysis of NMR data has been proposed, and some crucial steps in the pipeline are being performed manually, which constitutes a bottleneck in the experimental workflow, and takes weeks to months, depending on complexity of investigated protein. An automation of this process could significantly accelerate research in multiple domains,  including structure solving, protein-ligand interactions and structure-based drug discovery. 

A routine for NMR data analysis usually involves three consecutive steps: (a) peak picking, (b) chemical shift assignment, and (c) structure calculation.    

Peak picking consists in selecting signals (true peaks) in a NMR spectrum that come from protein atoms (see Fig. 1b). Usually, true peaks are characterized by specific regularities in shape, and can be visually distinguished from noise and other artifacts (signals arising due to imperfection of measurement system). The number of true peaks in a single spectrum varies from hundreds to thousands, depending on experiment type and protein complexity. 

Each true peak is a result of an interaction among a triplet or a quadruplet (depending on the dimensionality of a spectrum) of atoms during the NMR examination. Peak coordinates correspond to resonant frequencies of atoms in the tuple. Chemical shift assignment is a process of assigning atom identifiers to peaks, using intricate combination of various constraints and domain knowledge. 

Finally, having paired atom identifiers with their chemical shifts, we can use them as input to a method for protein structure calculation e.g. CYANA \cite{guntert04}. 

The scientific community has been struggling for 30 years to develop robust methods to support NMR data analysis. So far multiple approaches have been proposed for peak picking, including \cite{skinner2016,wurz2017}. One of the recent surveys lists 44 methods performing chemical shift assignment \cite{guerry2011}.

\begin{figure*}[h]
	\begin{minipage}[l]{0.75\textwidth}
	\includegraphics[width=\textwidth]{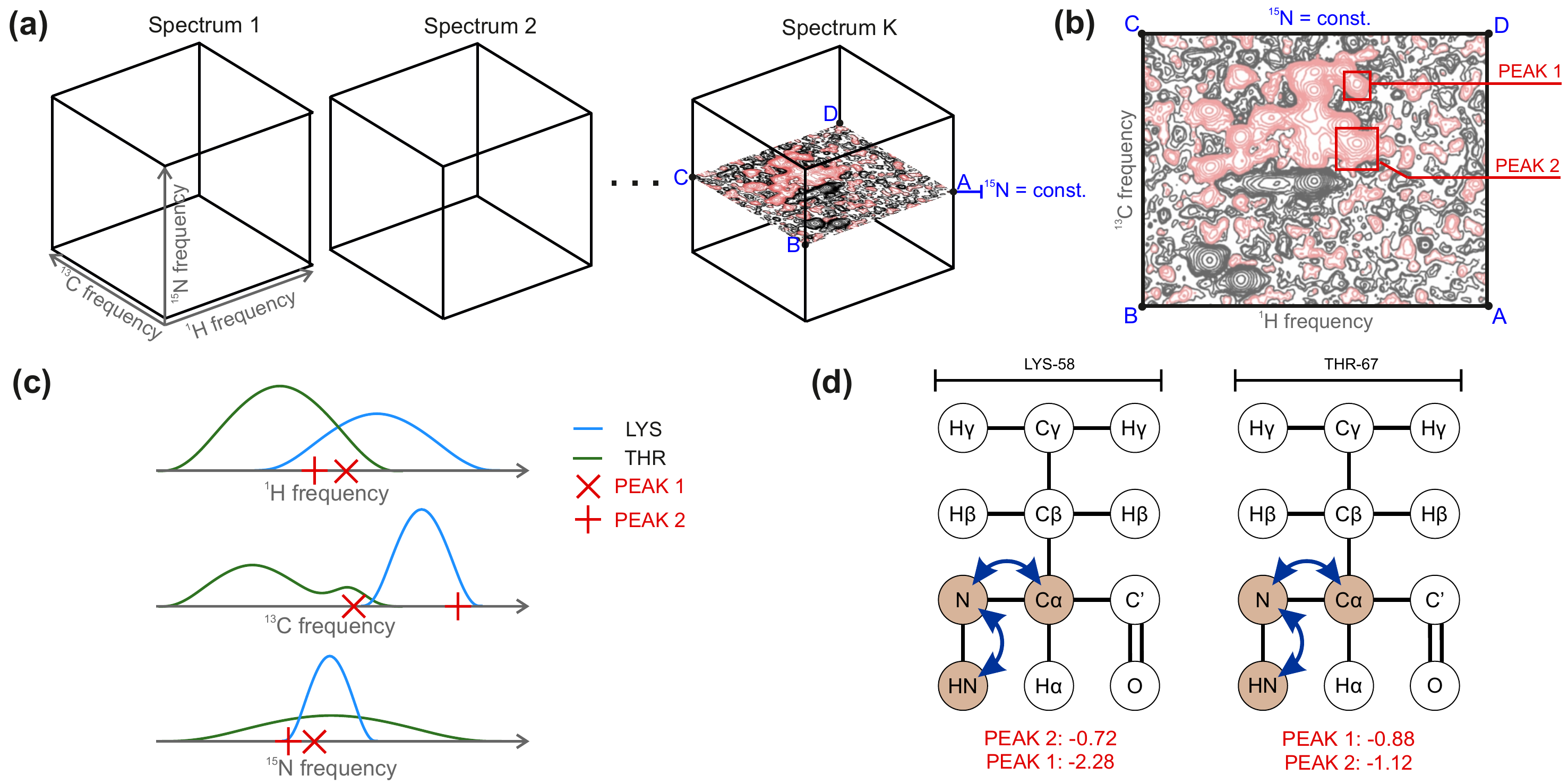}
    \end{minipage}
    \hspace{1mm}
	\begin{minipage}[l]{0.23\textwidth}
	\caption{(a) A set of multidimensional NMR spectra, recorded for a single protein. (b) A section through a spectrum. Black and red contour lines present positive and negative amplitude of a signal. Examples of true peaks are marked with frames. (c) Distributions of resonant frequencies for two different amino acids - lysine (LYS) and threonine (THR). (d) An exemplary matching of two peaks with two tuples of atoms. Scores are given under the amino acids.}
    \end{minipage}
\end{figure*}

Here, we present our ongoing research towards developing fully automated tool for chemical shift assignment, starting from a set of raw spectra at the input. We combine various machine learning techniques, including Convolutional Neural Networks (CNN) and Dirichlet Process Mixture Models (DPMM), with combinatorial optimization methods. So far, we have achieved state-of-the-art results on two intermediate steps of the analysis -  peak picking \cite{klukowski2018} and \textit{spin system identification} \cite{klukowski2018b}.

\section{Methods and Results}

Having a set of different spectra $S_{1},\ldots,S_{K}$ (Fig. 1a) measured for one protein, which contain information about resonant frequencies encoded in  peaks, the goal of chemical shift assignment is to determine a value of resonant frequency $z_{n}$ for each atom $n=1,\ldots,N$ in a protein sequence. Initially, for each spectrum $S_{k}$ we define tuples $\mathbf{z}_{j}^{(k)}$ of chemical shifts that encodes coordinates where theoretically should appear a true peak, and for each tuple we determine the types of amino acids $\mathbf{c}_{j}^{(k)}$ that the tuple is associated with. We can do so since the types of the spectra as well as protein sequence are known. However, notice that a particular chemical shift $z_{n}$ can appear in many tuples associated with the same or different spectra.
%
%

In our approach, we first utilize deep learning-based detection model that for each spectrum $S_{k}$ returns a list of peaks $\mathbf{e}_{i}^{(k)}$ together with the scores $\pi_{i}^{(k)}$, which are estimates of log probabilities of being a true peak (Fig 1b). We tested our method against other peak pickers, namely CV-Peak picker, NMRViewJ and CCPN, and achieved the mean increase in precision of 0.10, 0.15, and 0.27, respectively. A detailed description of our peak picking algorithm and more comprehensive experimental studies can be found in \cite{klukowski2018}.

Second, we estimate log probabilities $\xi_{ij}^{(k)} = \log p(\mathbf{e}_{i}^{(k)}|\mathbf{c}_{j}^{(k)})$ 
%
%
of occurrence of a peak assuming that it is associated with atoms belonging to given types of amino acids (Fig. 1c). Determining the form of $p(\mathbf{e}_{i}^{(k)}|\mathbf{c}_{j}^{(k)})$ 
%
%
is known as spin system identification, and here we utilize a non-parametric approach based on DPMM. We used over 10,000 proteins stored in BMRB database to train the model. We tested our approach on 36 experimental cases, and it outperformed reference method, based on kernel density estimation, in 31 out of 36 test cases. The advantage in accuracy varied from 6.48\% to 12.96\%. More details on the method is described in \cite{klukowski2018b}.

Next, for each spectrum we define a matrix $\mathbf{A}^{(k)}$ storing scores $a_{ij}^{(k)}=\pi_{i}^{(k)}+\xi_{ij}^{(k)}$ that combine observation evidence determined by the detection model with the spin system knowledge captured by DPMM. Finally, we define a set of matrices $\mathcal{X} = \{ \mathbf{X}^{(1)}, \mathbf{X}^{(2)}, ..., \mathbf{X}^{(K)} \}$, where $x_{ij}^{(k)} \in \{0, 1\}$ indicates if $i$-th peak in spectrum $k$ is associated with $j$-th tuple of atoms (Fig. 1d). Putting all together, we define chemical shift assignment as the following Mixed Integer Linear Program (MILP) with Indicator Constraints:

\begin{equation}
\arg \max_{\mathcal{X},\mathbf{z}} \sum_k \mathrm{tr} \left( \mathbf{A}^{(k)\top} \mathbf{X}^{(k)} \right)
\end{equation}
s.t.
\begin{equation}
\forall_k \forall_i \sum_j x_{ij}^{(k)} \leq 1 \quad \forall_k \forall_j \sum_i x_{ij}^{(k)} \leq 1
\end{equation}
\begin{equation}
\forall_k \forall_i \forall_j\ x_{ij}^{(k)}=1 \implies \|\mathbf{z}_{j}^{(k)}-\mathbf{e}_{i}^{(k)} \|\leq \varepsilon
\end{equation}
where $\varepsilon$ is a reasonably small value.
%

Due to high dimensionality of the problem in typical settings (the number of binary variables ranges from 10k to 5M for medium size proteins), it is intractable to use standard MILP optimizers. Therefore, we follow a homotopy-based approach, where we iteratively change optimization problem using constraints relaxation, improving lower bound in each iteration. 

For empirical evaluation we used ground truth data from BMRB records to generate true peak positions for selected proteins and spectra types. Surprisingly, we were able to find global maxima for protein sequences longer than 100 amino acids. Although the method requires more comprehensive studies, the initial results show that it is a promising direction towards fully automated chemical shift assignment. 




\bibliographystyle{named}
{
\small
\bibliography{ijcai16}

\begin{thebibliography}{}

\bibitem[\protect\citeauthoryear{Guerry and Herrmann}{2011}]{guerry2011}
Paul Guerry and Torsten Herrmann.
\newblock Advances in automated {NMR} protein structure determination.
\newblock {\em Quarterly reviews of biophysics}, 44(3):257--309, 2011.

\bibitem[\protect\citeauthoryear{G{\"u}ntert}{2004}]{guntert04}
Peter G{\"u}ntert.
\newblock Automated {NMR} structure calculation with {CYANA}.
\newblock {\em Protein NMR Techniques}, pages 353--378, 2004.

\bibitem[\protect\citeauthoryear{Klukowski \bgroup \em et al.\egroup
  }{2018a}]{klukowski2018b}
P~Klukowski, M~Augoff, M~Zamorski, A~Gonczarek, and MJ~Walczak.
\newblock Application of {D}irichlet process mixture model to the
  identification of spin systems in protein {NMR} spectra.
\newblock {\em Journal of biomolecular NMR}, 2018.

\bibitem[\protect\citeauthoryear{Klukowski \bgroup \em et al.\egroup
  }{2018b}]{klukowski2018}
Piotr Klukowski, Michal Augoff, Maciej Zieba, Maciej Drwal, Adam Gonczarek, and
  Michal~J. Walczak.
\newblock {NMRN}et: A deep learning approach to automated peak picking of
  protein {NMR} spectra.
\newblock {\em Bioinformatics}, 2018.

\bibitem[\protect\citeauthoryear{Skinner \bgroup \em et al.\egroup
  }{2016}]{skinner2016}
Simon~P Skinner, Rasmus~H Fogh, Wayne Boucher, Timothy~J Ragan, Luca~G Mureddu,
  and Geerten~W Vuister.
\newblock Ccpnmr analysisassign: a flexible platform for integrated {NMR}
  analysis.
\newblock {\em Journal of biomolecular NMR}, 66(2):111--124, 2016.

\bibitem[\protect\citeauthoryear{W{\"u}rz and G{\"u}ntert}{2017}]{wurz2017}
Julia~M W{\"u}rz and Peter G{\"u}ntert.
\newblock Peak picking multidimensional {NMR} spectra with the contour geometry
  based algorithm {CYPICK}.
\newblock {\em Journal of biomolecular NMR}, 67(1):63--76, 2017.

\end{thebibliography}
}

\end{document}